\documentclass[11pt]{article}
\usepackage{amsfonts,amsmath}
\title{Duality, monotonocity and the Wigner--Yanase--Dyson metrics}
\author{M. R. Grasselli \thanks{Research supported by the Natural Sciences and Engineering 
Research Council of Canada} \\ Dept. of Mathematics and Statistics \\
McMaster University \\ Hamilton ON L8S 4K1 \\ Canada}
\newtheorem{theorem}{Theorem}[section]

\newtheorem{lemma}[theorem]{Lemma}
\newtheorem{definition}{Definition}

\newcommand{\Man}{{\cal M}}
\newcommand{\Stat}{{\cal S}}
\newcommand{\Hilb}{{\cal H}}
\newcommand{\Alg}{{\cal A}}

\newcommand{\nalpha}{\nabla^{(\alpha)}}
\newcommand{\nmalpha}{\nabla^{(-\alpha)}}

\begin{document}
\maketitle
\begin{abstract}
We show that, for each value of $\alpha \in (-1,1)$, the only Riemannian
metrics on the space of positive definite matrices for which the 
$\nalpha$ and $\nmalpha$
connections are mutually dual are matrix multiples of the 
Wigner-Yanase-Dyson metric. If we further impose that the metric be monotone, then
this set is reduced to scalar multiples of the Wigner-Yanase-Dyson metric.
\end{abstract}

\section{Introduction}

Classical information geometry addresses the differential geometric properties
of families of classical probability densities. Quantum information geometry is its
noncommutative counterpart, dealing with the geometric structure of families of
quantum probabilities. The classical theory has been already explored and 
extended substantially, to the point of treating the geometric structures
of the infinite dimensional Banach manifold of all probability measures equivalent
to a given one \cite{PistoneSempi95,GibiliscoPistone98}. All the ingredients
of the original Amari's theory \cite{Amari85,AmariNagaoka00}, such as the Fisher
metric, the exponential, mixture and $\alpha$-connections, have been defined for 
this general manifold, from which the finite dimensional results follow by restricting
them to its finite dimensional submanifolds \cite{Grasselli01b}. In comparison, 
the quantum version still has ``miles to go before sleep'' \cite{Frost23}, 
being so far mostly restricted to the geometry of density matrices on 
finite dimensional Hilbert spaces. It stands as a proof of the richness of the quantum
domain that even this limited setup already offers many challenging problems, completely
absent in the classical case.

A central theme in the passage from classical to quantum information geometry is the
breakdown of Chentsov's result \cite{Chentsov82} that the Fisher metric  is the unique
 Riemannian metric (up to scalar multiples) on finite dimensional classical information
manifolds which is reduced by all Markov morphisms. As proved by Petz \cite{Petz96}, 
there are infinitely many Riemannian metrics on a matrix space with the property
of being reduced by stochastic maps (the quantum analogue of Markov morphisms). 
Having characterized all these possible
monotone metrics in terms of operator monotone functions, Petz's result opened the way 
to two different trends: to deal with the whole set 
of monotone metrics at once and try to find yet other characterizations 
\cite{LesniewskiRuskai99,GibiliscoIsola01,Petz02} or to find out which among them are more natural
then the others according to properties beyond monotonicity \cite{HasegawaPetz97,Uhlmann93}. 
This paper is dedicated to the second of these trends.
Its general attitude could be rephrase as: if monotonicity
is not enough to single out one particular metric, what are
the other conditions that should be further imposed in order to obtain a unique metric
on the information manifolds of density matrices ? The answer we offer is
based on the concept of duality for affine connections with respect to a given metric.  

There are two flat connections that can be introduced on information manifolds in a
fundamental way: the mixture connection, coming from the linear structure of the
manifold itself (either as a subset of $L^1$ in the classical case or as a subset of 
the trace class operators in the quantum case), and the exponential connection, 
coming from
the linear structure of their logarithms. The former, denoted by $\nabla^{(-1)}$
or $\nabla^{(m)}$, arises naturally when we consider
mixed states (classical or quantum), whereas the latter, denoted by $\nabla^{(1)}$
or $\nabla^{(e)}$, is intimately related to the 
concepts of moment generating functionals and partition functions. 
For infinite dimensional classical
manifolds, the exponential connection was rigorously defined in 
\cite{GibiliscoPistone98}, 
making use of exponential Orlicz spaces, while the mixture connection is similarly 
defined in \cite{Grasselli02a}, based on the conjugate Orlicz space of type $L\log L$. 
Of course the nonparametric definitions are designed in such a way that when restricted 
to finite dimensional submanifolds
they reduce to the long standing definitions of the parametric theory  
\cite{AmariNagaoka00}. For infinite dimensional quantum information manifolds, the
exponential connection was obtained in \cite{Streater00a,Streater00b,GrasselliStreater00}, 
using 
the technique of small perturbations of forms and operators in Hilbert spaces, but
the mixture connection poses a much harder problem, which is to some extent still open
\cite{Streater02}. Fortunately, the situation is straightforward as far as
finite dimensional quantum systems are concerned. Many authors have proposed 
essentially equivalent definitions for the exponential and mixture connections
on manifolds of density matrices \cite{Hasegawa95,Nagaoka95,Jencova01b}. We summarize our
views on these definitions for $\nabla^{(1)}$ and $\nabla^{(-1)}$ in 
\cite{GrasselliStreater01a}, where we observed that they are
flat connections by 
explicitly constructing affine coordinate systems for each of them. 

Two connections are said to be dual with respect to a metric if the combined action
of their parallel transport is compatible with the metric (see section 3 below
for the technical definition). The same pair of connections can be dual with respect to 
a multitude of metrics. It is then meaningful to ask, for a given pair of connections,
what are the all the possible metrics that make them dual. When we looked at the 
mixture and the exponential connection on finite dimensional quantum systems, 
we found in
\cite{GrasselliStreater01a} that the only metrics with this duality property are 
{\em matrix}
multiples of the Bogoliubov-Kubo-Mori inner product. Using Petz's characterization,
we then obtained the improved result that the only monotone metrics which
make the $\pm 1$-connections dual are {\em scalar} multiples of the {\em BKM} metric. The
purpose of the present paper is to investigate the same kind of question
for the more general pairs of $\pm\alpha$-connections.

In the classical version of Information Geometry, there are
two equivalent ways of defining the $\alpha$-connections $\nalpha$ on 
an information manifold ${\cal M}$, for $\alpha\in(0,1)$. 
The first approach consists of using the 
$\alpha$-embeddings of the form $p\mapsto \frac{2}{1-\alpha}p^{\frac{1-\alpha}{2}}$
to map  ${\cal M}$  into the sphere of radius $r$ in the Banach space $L^r$, for 
$r=\frac{2}{1-\alpha}$. One then looks at the natural connection on $L^r$, that is, 
the one for which the parallel transport is just the identity map, and its 
canonical projection onto the sphere of radius $r$. The pullback of the latter (again
using the $\alpha$-embedding) is then defined to be the $\alpha$-connection 
on ${\cal M}$. For finite dimensional manifolds, this can be traced back to the early
works
of Amari \cite{Amari85} and \cite{Chentsov82}, where they are 
introduced without explicitly mention of what the
target spaces for the $\alpha$-embeddings should be. For infinite dimensional 
information
manifolds, one has to explicitly make use of the functional analytic properties of the
spaces $L^r$ (namely that they are locally convex spaces), in order to unequivocally
define what is meant by the {\em canonical} projection onto a sphere. This was done in 
detail for the first time in \cite{GibiliscoPistone98} and in a slightly different fashion
in \cite{Grasselli02a}. In any event, one can prove that
\begin{equation}
\label{convex}
\nalpha=\frac{1+\alpha}{2}\nabla^{(1)}+\frac{1-\alpha}{2}\nabla^{(-1)},
\end{equation} 
which can then be taken as an equivalent definition for $\nalpha$. Proposals
for the quantum analogues of $\alpha$-connections, for both finite and infinite
dimensional manifolds, have appeared in number of papers 
\cite{Hasegawa97,Jencova01b,GibiliscoIsola99}. They all use the $\alpha$-embeddings in one way
or another. We present them in section 2, where we review some of their most
relevant properties. As it turns out, the $\alpha$-embedding definitions are no longer 
equivalent to (\ref{convex}), that is, to the definition based on the
convex mixture of the $\pm 1$-connections. We shall have more to 
say about this point later on in the paper.   

As it is well known, the {\em BKM} metric is a limiting case of the more general family
of Wigner-Yanase-Dyson metrics, denoted by $g^{\alpha}$ (more about this notation
later). 
The {\em WYD} metrics made their first appearance in the
context of quantum information geometry in the work of Hasegawa \cite{Hasegawa93}. 
It was later proved that they are monotone for all values of $\alpha\in [-3,3]$ \cite{HasegawaPetz96}.
In the spirit of the $\alpha$-embeddings discussed above, for which the target 
spaces are $L^r$, with $r=\frac{2}{1-\alpha}$, we restrict our discussion to the range
$\alpha\in (-1,1)$, thus corresponding to $r\in (1,\infty)$. It is straightforward
to prove
that, for each fixed value of $\alpha$ in this range, the $\pm\alpha$-connections
are dual with respect to the metric $g^{\alpha}$ \cite{Hasegawa95}. The formal limits 
$\alpha\rightarrow\pm 1$ lead to the {\em BKM} metric and the exponential and mixture
connections, for which the duality is established separately \cite{Nagaoka95}.

Following the same technique of \cite{GrasselliStreater01a}, we obtain the converse of
this result. We find in section 3 that,
for each fixed value of $\alpha\in (-1,1)$, the only metrics for which 
$\nalpha$ and $\nmalpha$ are dual are {\em matrix} multiples
of the {\em WYD} metric $g^{\alpha}$. Using Petz's characterization, we obtain
in section 4 that the only monotone metrics on positive definite matrices which
make the $\pm\alpha$-connections dual are {\em scalar} multiples of
$g^{\alpha}$.

\section{The quantum $\alpha$-connections} 
\label{alphaconn}

\subsection{The $\alpha$-representation}
\label{alpharepresentation}

Following the notation in \cite{GrasselliStreater01a}, let 
${\cal H}^N$ be a finite dimensional complex Hilbert space, 
${\cal B}({\cal H}^N)$ the algebra of operators on $\Hilb^N$, $\cal A$
its $N^2$-dimensional real vector subspace of self-adjoint operators and 
$\cal M$ the $n$-dimensional submanifold of all
invertible density operators on ${\cal H}^N$, with $n=N^2-1$. For
$\alpha\in (-1,1)$, define
the $\alpha$-embedding of $\cal M$ into $\cal A$ as
\begin{eqnarray*}
\ell_{\alpha} & : & {\cal M} \rightarrow{\cal A} \\
& & \rho \mapsto \frac{2}{1-\alpha}\rho^{\frac{1-\alpha}{2}}.
\end{eqnarray*}
Since $\cal A$ is itself a vector space, its tangent vectors consist of the partial
derivatives of curves in $\Alg$. Therefore we can use the $\alpha$-embedding
to obtain an explicit representation of the tangent bundle of $\cal M$ in terms of 
operators in $\cal A$, provided we can efficiently take partial derivatives
of functions of operators in $\Alg$. The noncommutative nature of quantum manifolds
makes a full appearance at this point, since the derivative of a matrix with respect
its parameters does not necessarily commute with the original matrix.
As a result, tools such as the chain rule do not hold in matrix 
calculus. To overcome this difficulty, at least for functions of
density matrices, we make use of the following decomposition. In the sequel, for 
$A\in{\cal B}({\cal H}^N)$, let ${\cal C}(A)=\{B\in{\cal B}({\cal H}^N):[A,B]=0\}$
denote its commutant.

\begin{lemma}[Hasegawa, 1997] Let $\Stat={\rho(\theta)}$ be a smooth manifold 
of invertible density matrices. Then there exist a anti-selfadjoint operator
$\Delta_i$ such that
\begin{equation}
\label{decomp1}
\frac{\partial\rho}{\partial\theta^i}=\frac{\partial^{c}\rho}{\partial\theta^i}
+[\rho,\Delta_i],\qquad \frac{\partial^{c}\rho}{\partial\theta^i}\in{\cal C}(\rho),
\quad [\rho,\Delta_i]\in{\cal C}(\rho)^\perp,
\end{equation}
the orthogonality being with respect to the Hilbert-Schmidt inner
product in ${\cal B}({\cal H}^N)$. Moreover, for any function F 
which is differentiable on a neighbourhood of the spectrum of $\rho$ we have
\begin{equation}
\label{decomp2}
\frac{\partial F(\rho)}{\partial\theta^i}=\frac{\partial^{c}F(\rho)}{\partial\theta^i}
+[F(\rho),\Delta_i],
\quad \frac{\partial^{c}F(\rho)}{\partial\theta^i}\in{\cal C}(\rho), 
\quad [F(\rho),\Delta_i]\in{\cal C}(\rho)^\perp.
\end{equation}
\label{decomp}
\end{lemma}

At each point $\rho \in {\cal M}$,
consider the subspace of $\cal A$ defined by
\[{\cal A}^{(\alpha)}_{\rho}=\left\{A \in {\cal A} : \mbox{Tr} 
\left(\rho^{\frac{1+\alpha}{2}} A\right) =0 \right\}.\]
Using (\ref{decomp2}) with
$F(\rho)=\ell_\alpha(\rho)$, we obtain
\begin{equation}
\label{alpharep}
\frac{\partial \ell_{\alpha}(\rho)}{\partial\theta^i}=
\rho^{\frac{1-\alpha}{2}}\frac{\partial^{c}\log \rho}{\partial\theta^i}
+\frac{2}{1-\alpha}[\rho^{\frac{1-\alpha}{2}},\Delta_i].
\end{equation}
Therefore, it follows from the normalization condition $\mbox{Tr}\rho=1$
and the cyclicity of the trace that
\[\mbox{Tr}\left(\rho^{\frac{1+\alpha}{2}}\frac{\partial \ell_\alpha(\rho)}
{\partial\theta^i}\right) = \mbox{Tr}\left(\frac{\partial^{c}\rho}{\partial\theta^i}
+\frac{2}{1-\alpha}[\rho,\Delta_i]\right)=0,\]
so that $\frac{\partial \ell_{\alpha}(\rho)}{\partial\theta^i}\in 
{\cal A}^{(\alpha)}_{\rho}$.

We can then define the isomorphism
\begin{eqnarray}
\label{alphaiso}
(\ell_\alpha)_{*(\rho)} & : & T_{\rho}{\cal M} \rightarrow {\cal A}^{(\alpha)}_{\rho}
\nonumber  \\
& & v \mapsto (\ell_{\alpha} \circ \gamma)^\prime (0),
\end{eqnarray}
where $\gamma :(-\varepsilon,\varepsilon) \rightarrow {\cal M}$ is a
curve in the equivalence class of the tangent vector $v$. We call this
isomorphism the
$\alpha$-representation of the tangent space $T_{\rho}{\cal M}$. If
$(\theta^1,\ldots ,\theta^n)$ is a coordinate system for $\cal M$,
then the $\alpha$-representation of the basis $\left\{\frac{\partial}{\partial\theta^1},
\ldots ,\frac{\partial}{\partial\theta^n}\right\}$ of $T_{\rho}{\cal M}$ is
$\left\{ \frac{\partial\ell_{\alpha}(\rho)}{\partial\theta^1},
\ldots ,\frac{\partial\ell_{\alpha}(\rho)}{\partial\theta^n}\right\}$. The
$\alpha$-representation of a vector field $X$ on $\cal M$ is therefore
the $\cal A$-valued function $(X)^{(\alpha)}$ given by
$(X)^{(\alpha)}(\rho)=(\ell_{\alpha})_{*(\rho)}X_{\rho}$.

\subsection{The covariant derivative $\nalpha$} 

The $\pm 1$-connections have a simple definition in terms of their parallel
transports, essentially because the $\pm 1$-embeddings map $\Man$ into
sets with an affine structure (the density operators themselves in the
$-1$-embedding and their logarithms in the $1$-embedding). Once their (flat)
parallel transports are defined, it is then a simple matter to find the coefficients
of their covariant derivatives, as well as to exhibit affine coordinate systems for
them, as explained for instance in the second section of \cite{GrasselliStreater01a}. 
However, as noted in the introduction, the $\alpha$-embeddings can be
viewed as a map from $\Man$ into the positive orthant of the sphere of radius 
$r=\frac{2}{1-\alpha}$ in 
$\Alg$ when we equip $\Alg$ with the the $r$-norm
\[\|A\|_r:=\left(\mbox{Tr}|A|^r\right)^{1/r}.\]
Indeed, we can readily verify that, for any $\rho\in\Man$,
\[\|\ell_{\alpha}(\rho)\|_r=\left(\mbox{Tr}\left|r\rho^{1/r}\right|^r\right)
^{1/r}=r,\]
so that $\ell_{\alpha}(\rho)\in S^r$, the sphere of radius $r$ in $\Alg$. More
interestingly, it can be shown that the tangent space at a point 
$0\leq\sigma\in S^r$ is
\[T_{\sigma}S^r=\left\{A\in\Alg:\mbox{Tr}(A\sigma^{r-1})=0\right\}\]
(see the second section of \cite{GibiliscoIsola99} 
for a quick review of the geometry of spheres in the more general context of 
uniformly convex Banach spaces).
If we put $\sigma=\ell_{\alpha}(\rho)=r\rho^{1/r}$, we find that
\[T_{r\rho^{1/r}}S^r=\left\{A\in\Alg:\mbox{Tr}(A\rho^{1-1/r})=0\right\}=
{\cal A}^{(\alpha)}_{\rho},\]
so that the $\alpha$-representation (\ref{alphaiso}) is indeed an isomorphism
between tangent spaces, as the push-forward notation suggests.

The sphere $S^r$ inherits a natural connection obtained by projecting the trivial 
connection on $\Alg$ (the one where parallel transport is just the identity map)
onto its tangent space at each point. For each $0\leq\sigma\in S^r$, the
canonical projection from the tangent space $T_{\sigma}\Alg$ onto
the tangent space $T_{\sigma}S^r$ is uniquely given by \cite{GibiliscoIsola99}
\begin{eqnarray*}
\Pi_{\sigma} & : & T_{\sigma}\Alg \rightarrow T_{\sigma}S^r 
\\  &   & A \mapsto A-\left(r^{-r}\mbox{Tr}\left[A\sigma^{r-1}\right]\right)\sigma.
\end{eqnarray*}
For $\sigma=\ell_{\alpha}(\rho)=r\rho^{1/r}$, this gives
\begin{eqnarray*}
\Pi_{r\rho^{1/r}} & : & T_{r\rho^{1/r}}\Alg \rightarrow T_{r\rho^{1/r}}S^r 
\\  &   & A \mapsto A-\left(\mbox{Tr}\left[\rho^{\frac{1+\alpha}{2}}A\right]\right)
\rho^{\frac{1-\alpha}{2}}.
\end{eqnarray*}

We can now define the covariant derivative of the $\alpha$-connection. Starting
with a differentiable vector field $s\in S(T\Man)$, we first push it forward under
the $\alpha$-embedding along a curve $\gamma$ to obtain 
$(\ell_\alpha)_{*(\gamma(t))}s \in T\Alg$. We then take its covariant derivative
with respect to the trivial connection on $\Alg$, denoted by $\widetilde{\nabla}$,
in the direction of $(\ell_\alpha)_{*(\rho)}v$, that is, the push-forward of a
tangent vector $v\in T_{\rho}\Man$. The result is a vector in 
$T_{r\rho^{1/r}}\Alg$, which we then project down to $T_{r\rho^{1/r}}S^r$ using
the operator $\Pi_{r\rho^{1/r}}$ above. Finally, we pull it back to 
$T_{\rho}\Man$ using $(\ell_\alpha)^{-1}_{*(\rho)}$ and call it the 
$\alpha$-covariant derivative of the vector field $s$ in the direction of the 
tangent vector $v$ at the point $\rho\in\Man$. 
The formula for all these operations reads like the
following.

\begin{definition} For $\alpha \in (-1,1)$, let $\gamma :
(-\varepsilon ,\varepsilon)\rightarrow
{\cal M}$ be a smooth curve such that $\rho=\gamma (0)$ and $v=\dot{\gamma}(0)$ 
and let $s \in S(T{\cal M})$ be a differentiable vector field. The 
$\alpha$-connection on $T{\cal M}$ is given by
\begin{equation}
\left(\nabla^{(\alpha)}_vs\right)(\rho)=(\ell_{\alpha})_{*(\rho)}^{-1}
\left[\Pi_{r\rho^{1/r}}\widetilde{\nabla}_{(\ell_{\alpha})_{*(\rho)}v}
(\ell_{\alpha})_{*(\gamma(t))}s\right].
\label{alpha}
\end{equation}
\label{def_alpha}
\end{definition}

Using the definition (\ref{alpha}), we find that the $\alpha$-representation
of the $\alpha$-covariant derivative of the vector field 
$\partial/\partial\theta^j$ in the direction of the tangent vector
$\partial_i:=\partial/\partial\theta^i$ is
\begin{equation}
\label{alpha-cov}
\left(\nabla^{(\alpha)}_{\partial_i}\frac{\partial}
{\partial\theta^j}\right)^{(\alpha)}=
\frac{\partial^2\ell_{\alpha}(\rho)}{\partial\theta^i\partial\theta^j}-
\mbox{Tr}\left(\rho^{\frac{1+\alpha}{2}}\frac{\partial^2\ell_{\alpha}(\rho)}
{\partial\theta^i\partial\theta^j}\right)\rho^{\frac{1-\alpha}{2}}.
\end{equation}

\subsection{The $\alpha$-parallel transport and the extend manifold $\widehat{\Man}$}  

The $\alpha$-parallel transport of a tangent vector from tangent spaces at
different points in $\Man$ is the pull-back of the parallel transport of its
$\alpha$-representation in $\Alg$. The latter, by its turn, consists of identity map 
followed by the canonical projection onto the $TS^r$ at all points along
a curve on $S^r$. It is obviously path dependent, and therefore no longer flat, like
the $\pm 1$-parallel transports were. This is a consequence of the fact that
among all $L^p$-spaces, for $1\leq p \leq\infty$, only the spaces $L^1$ and
$L^\infty$ have spheres which are flat with respect to their trivial connections (recall
the shape of the unit circles in $\mathbb{R}^2$ for all the different $L^p$-norms). 

Now let us consider the extended manifold of faithful
weights $\widehat{\Man}$ (the positive definite matrices). Observe first that the
$\alpha$-embedding in this case maps $\widehat{\Man}$ to itself. Moreover, for any $\sigma\in\widehat{\cal M}$,
$T_\sigma\widehat{\Man}=T_\sigma\Alg\simeq\Alg$, so that there is no need to do any projection in order
to obtain the parallel transport on $\widehat{\Man}$ induced by the $\alpha$-embedding.
We can therefore define the $\alpha$-parallel transport on $\widehat{\Man}$ simply by
\begin{eqnarray*}
\widehat{\tau}^{(\alpha)}_{\sigma_0,\sigma_1} & : & T_{\sigma_0}\widehat{\cal M} 
\rightarrow T_{\sigma_1}\widehat{\cal M}
\\                   &   & v \mapsto
(\ell_{\alpha})_{*(\sigma_1)}^{-1}\left((\ell_{\alpha})_{*(\sigma_0)}v\right),
\end{eqnarray*} and we find (using (\ref{alpha}) without the projection step)
that the $\alpha$-representation of its covariant 
derivative is 
\begin{equation}
\label{extalpha-cov}
\left(\widehat{\nabla}^{(\alpha)}_{\partial_i}
\frac{\partial}{\partial\theta^j}\right)^{(\alpha)}=
\frac{\partial^2\ell_{\alpha}(\rho)}{\partial\theta^i\partial\theta^j},
\end{equation}
where $\theta=\{\theta^1,\ldots,\theta^{n+1}\}$ 
is any coordinate system
for the extended manifold $\widehat{\Man}$. 
Now let $\{X_1,\ldots,X_{n+1}\}$ be a basis for $\Alg$. For each $\sigma\in\widehat{\Man}$, 
we have that $\sigma^{\frac{1-\alpha}{2}}\in\Alg$, so that there exist real
numbers $\xi=\{\xi^1,\ldots,\xi^{n+1}\}$ such that
\[\frac{2}{1-\alpha}\sigma^{\frac{1-\alpha}{2}}=\xi^1X_1+\cdots
+\xi^{n+1}X_{n+1}.\]
Then $\xi=\{\xi^1,\ldots,\xi^{n+1}\}$ is a 
$\widehat{\nabla}^{(\alpha)}$-affine
coordinate system for $\widehat{\Man}$, since (\ref{extalpha-cov}) gives
\[
\left(\widehat{\nabla}^{(\alpha)}_{\partial_i}
\frac{\partial}{\partial\xi^j}\right)^{(\alpha)}=
\frac{\partial^2\ell_{\alpha}(\rho)}{\partial\xi^i\partial\xi^j}=
\frac{\partial X_j}{\partial\xi^i}=0.
\]

Therefore, $\widehat{\Man}$ is $\widehat{\nabla}^{(\alpha)}$-flat, even though its
submanifold $\Man$ is not $\nalpha$-flat. We note in passing that the 
connection $\nalpha$
on the submanifold $\Man$ is a restriction of the connection $\widehat{\nabla}^{(\alpha)}$,
which acts on the larger manifold $\widehat{\Man}$, obtained without the use of any 
metric on $\Man$, but rather using the canonical projection existing in $\Alg$, the
target space for the $\alpha$-embedding.  

We finish this section with a couple of comparative remarks. Definition 
\ref{def_alpha} is the verbatim analogue for finite dimensional quantum systems
of the general definition for $\alpha$-connections for infinite dimensional
classical information manifolds \cite{GibiliscoPistone98,Grasselli02a} and
are, consequently, the quantum analogue of the original definition by
Amari \cite{Amari85} and Chentsov \cite{Chentsov82} as well. Formulae (\ref{alpha-cov})
and (\ref{extalpha-cov}) are special cases of those obtained by Jen\v{c}ov\'{a} using an
embedding by a more general monotone function $g$, which include the
$\alpha$-embeddings (see respectively line 3, page 150 and line 10, page 149 of
\cite{Jencova01a}). Finally, quantum $\alpha$-connection in the spirit we present here
had been hinted before by Hasegawa in \cite[equation 35]{Hasegawa95} and 
\cite[equation 16]{Hasegawa97}, although in the less general form of Christoffel's 
symbols, which depend on a metric to be defined, as opposed to covariant 
derivatives and parallel transports, which are therefore more intrinsic. Infinite
dimensional quantum $\alpha$-connections were proposed in \cite{GibiliscoIsola99},
making heavy use of the geometry of uniformly convex Banach spaces, of which
the definitions given here are concrete finite dimensional realizations.

\section{Duality and the {\em WYD} metrics}

We recall some purely geometrical definitions of duality, which apply to 
any statistical manifold, classical or quantum: dual affine connections and 
dual coordinate systems. 

Two connections $\nabla$ and $\nabla^*$ on a 
Riemannian manifold $({\cal M},g)$
are dual with respect to $g$ if and only if
\begin{equation}
X g(Y,Z)= g \left(\nabla_X Y,Z \right)+g \left(Y,\nabla^*_X Z\right),
\end{equation}
for any vector fields $X,Y,Z$ on $\cal M$ \cite{Amari85,MurrayRice93}. Equivalently,
if $\tau_{\gamma(t)}$ and $\tau^*_{\gamma(t)}$ are the respective parallel
transports along a curve $\{\gamma(t)\}_{0\leq t\leq 1}$ on $\cal M$, with
$\gamma(0)=\rho$, then $\nabla$ and $\nabla^*$ are dual with respect to $g$
if and only if for all $t\in[0,1]$,
\begin{equation}
g_\rho(Y,Z)=g_{\gamma(t)}\left(\tau_{\gamma(t)}Y,\tau^*_{\gamma(t)}Z\right).
\end{equation}

Two coordinate systems $\theta = (\theta^i)$ and $\eta = (\eta_i)$ 
on a Riemannian manifold
$({\cal M},g)$ are dual with respect to $g$ if and only if their
natural bases for $T_\rho{\cal M}$ are {\em biorthogonal} at every point
$\rho\in {\cal M}$, that is,
\[g\left(\frac{\partial}{\partial\theta^i},\frac{\partial}{\partial\eta_j}\right)=\delta^i_j.\]
Equivalently, $\theta = (\theta^i)$ and $\eta = (\eta_i)$ are dual
with respect to $g$ if and only if
\[g_{ij}=\frac{\partial\eta_i}{\partial\theta^j} \quad \mbox{and}
\quad g^{ij}=\frac{\partial\theta_i}{\partial\eta^j}\]
at every point $\rho\in {\cal M}$, where, as usual,
$g^{ij}=(g_{ij})^{-1}$.

The next two theorems establishes the role of potential functions as well as 
the relation between dual connections and dual coordinate systems for the
case of flat manifolds. In the sense used in this paper, 
a connection $\nabla$ on manifold $\cal M$ is
said to be flat if $\cal M$ admits a global $\nabla$-affine
coordinate system. This is equivalent to its curvature and
torsion both being zero.

\begin{theorem}[Amari, 1985]
When a Riemannian manifold $({\cal M},g)$ has a pair of dual coordinate
systems $(\theta,\eta)$, there exist potential functions $\Psi(\theta)$
and $\Phi(\eta)$ such that
\[g_{ij}(\theta)=\frac{\partial^2\Psi(\theta)}{\partial\theta^i\partial\theta^j}
\quad \mbox{and} \quad
g^{ij}=\frac{\partial^2\Phi(\eta)}{\partial\eta_i\partial\eta_j}.\]
Conversely, when either potential function $\Psi$ or $\Phi$ exists
from which the metric is derived by differentiating it twice, there
exist a pair of dual coordinate systems. The dual
coordinate systems and the potential functions are related by the following Legendre
transforms
\[\theta^i=\frac{\partial\Phi(\eta)}{\partial\eta_i},
\quad \eta_i=\frac{\partial\Psi(\theta)}{\partial\theta^i}\]
and
\[\Psi(\theta)+\Phi(\eta)-\theta^i\eta_i = 0\]
\label{potentials}
\end{theorem}

\begin{theorem}[Amari, 1985]
Suppose that $\nabla$ and $\nabla^*$ are two flat connections on a manifold $\cal M$.
If they are dual with respect to a Riemannian metric $g$ on
$\cal M$, then there exists a pair $(\theta,\eta)$
of dual coordinate systems such that $\theta$ is $\nabla$-affine
and $\eta$ is a $\nabla^*$-affine.
\label{dualities}
\end{theorem}

Let us now consider the definition of a Riemannian metric for our
manifold $\cal M$ of density matrices. Using the $\alpha$-representation
to obtain a concrete realization of tangent vectors 
on $\Man$ in terms of operators in ${\cal A}$, 
a Riemannian metric on $\Man$ is deemed to be provided by the
smooth assignment of an inner product $\langle\cdot,\cdot\rangle_
{\rho}$ in
${\cal A}\subset B({\cal H}^N)$ for each point $\rho \in {\cal M}$.

For a fixed $\alpha\in (-1,1)$, the {\em WYD} (Wigner-Yanase-Dyson) 
metric on $\cal M$ is given by
\begin{equation}
\label{gwyd}
g^{(\alpha)}_{\rho}(A,B) :=  \mbox{Tr}\left(A^{(\alpha)}
B^{(-\alpha)}\right), \qquad A,B \in T_{\rho}\Man.
\end{equation}

The symmetry properties of this definition are more apparent if one express it in
a coordinate system $(\theta^1,\ldots ,\theta^n)$ for $\Man$. By virtue of the
decomposition lemma \ref{decomp}, we have that

\begin{eqnarray}
g^{(\alpha)}_{ij}(\theta)&:=& g^{(\alpha)}_{\rho}\left(\frac{\partial}
{\partial\theta^i},\frac{\partial}{\partial\theta^j}\right)
=\mbox{Tr}\left(\frac{\partial\ell_{\alpha}(\rho)}
{\partial\theta^i}
\frac{\partial\ell_{-\alpha}(\rho)}{\partial\theta^j}\right)  \\
&=& \mbox{Tr}\left(\rho\frac{\partial^c\log\rho}{\partial\theta^i}
\frac{\partial^c\log\rho}{\partial\theta^j}\right)+\frac{4}{1-\alpha^2}\mbox{Tr}
\left[\rho^{\frac{1-\alpha}{2}},\Delta_i\right]\left[\rho^{\frac{1+\alpha}{2}},
\Delta_j\right]. \nonumber
\end{eqnarray}

It is then clear that $g^{(\alpha)}_{ij}=g^{(\alpha)}_{ji}=g^{(-\alpha)}_{ij}$. Observe
also that for the extreme cases $\alpha\rightarrow \pm 1$, formula (\ref{gwyd}) leads
to the familiar {\em BKM} (Bogoliubov-Kubo-Mori) metric  
\begin{equation}
g^{(\pm 1)}_{\rho}(A,B)=
g^{\scriptscriptstyle B}_{\rho}(A,B)  =  \mbox{Tr}\left(A^{(-1)}
B^{(1)}\right)
\end{equation}
where $A^{(\pm 1)},B^{(\pm 1)}$ are the $\pm 1$-representations of the tangent vectors
$A,B \in T_{\rho}\Man$, as explained, for instance, in \cite{GrasselliStreater01a}. In 
coordinates, the {\em BKM} metric assumes the form
\begin{eqnarray}
 g^{\scriptscriptstyle B}_{ij}(\theta)&:=& g^{\scriptscriptstyle B}_{\rho}
\left(\frac{\partial}
{\partial\theta^i},\frac{\partial}{\partial\theta^j}\right)
=  \mbox{Tr}\left(\frac{\partial\log\rho}
{\partial\theta^i}
\frac{\partial\rho}{\partial\theta^j}\right) \nonumber \\
&=&  \mbox{Tr}\left(\rho\frac{\partial^c\log\rho}{\partial\theta^i}
\frac{\partial^c\log\rho}{\partial\theta^j}\right)+\mbox{Tr}
[\log\rho,\Delta_i][\rho,\Delta_j].
\end{eqnarray}

It follows directly from the definition (\ref{gwyd}), as has been observed 
in a number of papers \cite{Hasegawa97,Jencova01b}, that the 
$\pm\alpha$-connections are dual with respect to the metric $g^{(\alpha)}$ for each
fixed value of $\alpha\in(-1,1)$ (just as the $\pm 1$-connections are dual
with respect to the {\em BKM} metric). Our purpose is to discover what other
metrics have the same property. 

As suggested by the statement in theorem \ref{dualities}, most of the 
ingredients of Amari's theory, such
as statistical divergences and the projection theorems \cite[pp. 84-93]{Amari85},
can only be {\em a priori} defined for flat manifolds. Only in a later stage, one
consider what happens when they are applied to curved submanifolds of flat
manifolds. Following this trend, we from now on confine our attention to those 
metrics on $\Man$ which
are obtained as restrictions of metrics on the extended manifold $\widehat{\Man}$,
which is $\widehat{\nabla}^{(\pm\alpha)}$-flat, and treat the latter as
our primary objects  

Observe first that the {\em WYD} metric extends quite
naturally to $\widehat{\Man}$, simply using the $\pm\alpha$-representations of
tangent vectors $\widehat{A},\widehat{B}$ (that is, the representation
induced by the $\pm\alpha$-embedding of $\widehat{\Man}$ into $\Alg$):
\begin{equation}
\widehat{g}^{(\alpha)}_{\sigma}\left(\widehat{A},\widehat{B}\right):=
\mbox{Tr}\left(\widehat{A}^{(\alpha)}\widehat{B}^{(-\alpha)}\right), 
\qquad \widehat{A},\widehat{B} \in T_{\sigma}\widehat{\Man}.
\end{equation}
It is also obvious that $\widehat{g}^{(\alpha)}$ has the same symmetry and duality 
properties of $g^{(\alpha)}$. We now show how $\widehat{g}^{(\alpha)}$ can be obtained 
from a potential function on $\widehat{\Man}$.

\begin{lemma} If $(\theta^1,\ldots ,\theta^{n+1})$ is a 
$\widehat{\nabla}^{(\alpha)}$-affine coordinate system 
for the extended manifold $\widehat{\Man}$, then the function
\begin{equation}
\widetilde{\Psi}_{\alpha}(\theta)=\frac{2}{1+\alpha} Tr \sigma(\theta),
\qquad \sigma(\theta)\in\widehat{\Man}
\label{alphapot}
\end{equation}
satisfies
\begin{equation}
\widehat{g}^{(\alpha)}_{ij}(\theta)=\frac{\partial^2\widetilde{\Psi}_{\alpha}(\theta)}
{\partial\theta^i\partial\theta^j}.
\end{equation}
Moreover, 
\begin{equation}
\widetilde{\eta}_i=\frac{\partial\widetilde{\Psi}_{\alpha}(\theta)}{\partial\theta^i}
\end{equation}
is a $\widehat{\nabla}^{(-\alpha)}$-affine coordinate system for $\widehat{\Man}$.
\label{alphalemma}
\end{lemma}
{\em Proof:} Since $\theta$ is $\widehat{\nabla}^{(\alpha)}$-affine, there exist linearly
independent operators $\left\{X_1, \ldots, X_{n+1}\right\}$ such that
\begin{equation}
\ell_{\alpha}(\sigma)=\frac{2}{1-\alpha}\sigma^{\frac{1-\alpha}{2}}=\theta^1X_1+\cdots
+\theta^{n+1}X_{n+1}.
\end{equation}
Since the point $\sigma\in\widehat{\Man}$ is fixed in the course of this proof, we 
omit it from the notation and just write $\ell_\alpha$ and $\ell_{-\alpha}$ for
$\ell_{\alpha}(\sigma)$ and $\ell_{-\alpha}(\sigma)$, respectively. 
From lemma \ref{decomp} we obtain that
\begin{equation}
X_i=\frac{\partial\ell_{\alpha}}{\partial\theta^i}=
\frac{\partial^{c}\ell_{\alpha}}{\partial\theta^i}+[\ell_{\alpha},\Delta_i],
\end{equation}
that is
\begin{equation}
\frac{\partial^{c}\ell_{\alpha}}{\partial\theta^i}=X_i+[\Delta_i,\ell_{\alpha}].
\end{equation}
Also, since
\[\ell_{-\alpha}=\frac{2}{1+\alpha}\sigma^{\frac{1+\alpha}{2}}=
\left(\frac{2}{1+\alpha}\right)\left(\frac{1-\alpha}{2}\right)^{\frac{1+\alpha}
{1-\alpha}}\ell_{\alpha}^{\frac{1+\alpha}{1-\alpha}}\]
we have that
\[\frac{\partial^{c}\ell_{-\alpha}}{\partial\theta^j}
=\left(\frac{1-\alpha}{2}\right)^{\frac{2\alpha}{1-\alpha}}
\ell_{\alpha}^{\frac{2\alpha}{1-\alpha}}
\frac{\partial^{c}\ell_{\alpha}}{\partial\theta^j}.\]
So using lemma \ref{decomp} again we get
\begin{eqnarray}
\frac{\partial\ell_{-\alpha}}{\partial\theta^j}&=&
\frac{\partial^{c}\ell_{-\alpha}}{\partial\theta^j}
+[\ell_{-\alpha},\Delta_j] \nonumber \\
&=& \left(\frac{1-\alpha}{2}\right)^{\frac{2\alpha}{1-\alpha}}
\ell_{\alpha}^{\frac{2\alpha}{1-\alpha}}
\frac{\partial^{c}\ell_{\alpha}}{\partial\theta^j}+ 
[\ell_{-\alpha},\Delta_j] \nonumber \\
&=& \left(\frac{1-\alpha}{2}\right)^{\frac{2\alpha}{1-\alpha}}
\ell_{\alpha}^{\frac{2\alpha}{1-\alpha}}\left(X_j+
[\Delta_j,\ell_{\alpha}]\right)+[\ell_{-\alpha},\Delta_j].
\label{ellm}
\end{eqnarray}
Now observe that
\begin{eqnarray}
\frac{\partial^2\widetilde{\Psi}_{\alpha}(\theta)}
{\partial\theta^i\partial\theta^j}&=&\frac{\partial^2}{\partial\theta^i\partial\theta^j}
\left(\frac{2}{1+\alpha}\mbox{Tr}\sigma\right)=\frac{2}{1+\alpha}\mbox{Tr}
\left(\frac{\partial^2 \sigma}{\partial\theta^i\partial\theta^j}\right)\nonumber\\
&=&\frac{2}{1+\alpha}\mbox{Tr}\left(\frac{\partial^2 \sigma^{\frac{1-\alpha}{2}}
\sigma^{\frac{1+\alpha}{2}}}{\partial\theta^i\partial\theta^j}\right) =
\frac{1-\alpha}{2}\mbox{Tr}\left(\frac{\partial^2 \ell_{\alpha}\ell_{-\alpha}}
{\partial\theta^i\partial\theta^j}\right)\nonumber\\
&=&\frac{1-\alpha}{2}\mbox{Tr}\left[\frac{\partial}{\partial\theta^i}\left(
X_j\ell_{-\alpha}+\ell_{\alpha}\frac{\partial\ell_{-\alpha}}{\partial\theta^j}\right)
\right]\nonumber\\&=&\frac{1-\alpha}{2}\mbox{Tr}\left(
X_j\frac{\partial\ell_{-\alpha}}{\partial\theta^i}+X_i\frac{\partial\ell_{-\alpha}}
{\partial\theta^j}+\ell_{\alpha}\frac{\partial^2\ell_{-\alpha}}{\partial\theta^i
\partial\theta^j}\right).
\label{ronaldo}
\end{eqnarray}
Let us now evaluate each of the terms in the last expression separately. For the
first one we have
\begin{eqnarray}
\mbox{Tr}\left(X_j\frac{\partial\ell_{-\alpha}}{\partial\theta^i}\right)&=&
\mbox{Tr}\left(X_j\frac{\partial^{c}\ell_{-\alpha}}{\partial\theta^i}
+X_j[\ell_{-\alpha},\Delta_i]\right)\nonumber\\
&=& \mbox{Tr}\left(X_j\frac{\partial^{c}\ell_{-\alpha}}{\partial\theta^i}
+\ell_{-\alpha}[X_j,\Delta_i]\right)\nonumber\\
&=& \mbox{Tr}\left(X_j\frac{\partial^{c}\ell_{-\alpha}}{\partial\theta^i}
+[\Delta_j,\ell_{\alpha}]\frac{\partial^{c}\ell_{-\alpha}}{\partial\theta^i}\right)
\nonumber\\
&=& \mbox{Tr}\left(\frac{\partial^{c}\ell_{\alpha}}{\partial\theta^j}
\frac{\partial^{c}\ell_{-\alpha}}{\partial\theta^i}\right),
\label{juninho}
\end{eqnarray}
where we have used that facts that $[A,\Delta_i]=0$ for any constant 
(independent of $\theta^i$) operator $A$ and 
$\mbox{Tr}\left([\Delta_j,\ell_{\alpha}]
\frac{\partial^{c}\ell_{-\alpha}}{\partial\theta^i}\right)=0$, since 
$\frac{\partial^{c}\ell_{-\alpha}}{\partial\theta^i}$ commutes with $\ell_{\alpha}$.
Exchanging the roles of the indices $i$ and $j$ in (\ref{juninho}) 
we find that the second term in (\ref{ronaldo}) gives
\begin{equation}
\mbox{Tr}\left(X_i\frac{\partial\ell_{-\alpha}}{\partial\theta^j}\right)
= \mbox{Tr}\left(\frac{\partial^{c}\ell_{\alpha}}{\partial\theta^i}
\frac{\partial^{c}\ell_{-\alpha}}{\partial\theta^j}\right).
\end{equation}
But
\[\frac{\partial^{c}\ell_{\alpha}}{\partial\theta^i}
\frac{\partial^{c}\ell_{-\alpha}}{\partial\theta^j}
=\sigma\frac{\partial^{c}\log\sigma}{\partial\theta^i}
\frac{\partial^{c}\log\sigma}{\partial\theta^j}=
\frac{\partial^{c}\ell_{-\alpha}}{\partial\theta^i}
\frac{\partial^{c}\ell_{\alpha}}{\partial\theta^j}.\]
Therefore
\begin{equation}
\mbox{Tr}\left(X_i\frac{\partial\ell_{-\alpha}}{\partial\theta^j}\right)
= \mbox{Tr}\left(\frac{\partial^{c}\ell_{\alpha}}{\partial\theta^i}
\frac{\partial^{c}\ell_{-\alpha}}{\partial\theta^j}\right)=
\mbox{Tr}\left(X_j\frac{\partial\ell_{-\alpha}}{\partial\theta^i}\right).
\label{rivaldo}
\end{equation}

As for the third term in (\ref{ronaldo})
\begin{equation*}
\begin{split}
\mbox{Tr}&\left(\ell_{\alpha}\frac{\partial^2\ell_{-\alpha}}{\partial\theta^i
\partial\theta^j}\right)=
 \mbox{Tr}\left(\ell_{\alpha}\frac{\partial}{\partial\theta^i}
\left\{\frac{\partial\ell_{-\alpha}}
{\partial\theta^j}\right\}\right)\\
&=\mbox{Tr}\left(\ell_{\alpha}\frac{\partial}{\partial\theta^i}
\left\{\left(\frac{1-\alpha}{2}\right)^{\frac{2\alpha}{1-\alpha}}
\ell_{\alpha}^{\frac{2\alpha}{1-\alpha}}
\frac{\partial^{c}\ell_{\alpha}}{\partial\theta^j}+
[\ell_{-\alpha},\Delta_j]\right\}\right)\\
&=\mbox{Tr}\left\{\left(\frac{1-\alpha}{2}\right)^{\frac{2\alpha}{1-\alpha}}
\ell_{\alpha}\frac{\partial\ell_{\alpha}^{\frac{2\alpha}{1-\alpha}}}
{\partial\theta^i}
\frac{\partial^{c}\ell_{\alpha}}{\partial\theta^j}+\ell_{\alpha}
\frac{\partial[\ell_{-\alpha},\Delta_j]}{\partial\theta^i}\right.\\
&\phantom{blabla}\left.+\left(\frac{1-\alpha}{2}\right)^{\frac{2\alpha}
{1-\alpha}}
\ell_{\alpha}^{\frac{1+\alpha}{1-\alpha}}\frac{\partial}
{\partial\theta^i}\left(
X_j+[\Delta_j,\ell_{\alpha}]\right)\right\}.
\end{split}
\end{equation*}
Now we use lemma \ref{decomp} once more in
\[\frac{\partial\ell_{\alpha}^{\frac{2\alpha}{1-\alpha}}}{\partial
\theta^i}=\left(\frac{2\alpha}{1-\alpha}\right)\ell_{\alpha}^{\frac{2\alpha}
{1-\alpha}-1}\frac{\partial^{c}\ell_{\alpha}}{\partial\theta^i}
+[\ell_{\alpha}^{\frac{2\alpha}{1-\alpha}},\Delta_i],\]
which inserted back in the last equation gives
\begin{eqnarray}
\mbox{Tr}\left(\ell_{\alpha}\frac{\partial^2\ell_{-\alpha}}
{\partial\theta^i\partial\theta^j}\right)&=&
\mbox{Tr}\left\{\left(\frac{1-\alpha}{2}\right)^{\frac{2\alpha}{1-\alpha}}
\left(\frac{2\alpha}{1-\alpha}\ell_{\alpha}^{\frac{2\alpha}{1-\alpha}}
\frac{\partial^{c}\ell_{\alpha}}{\partial\theta^i}+[\ell_{\alpha}^
{\frac{1+\alpha}{1-\alpha}},\Delta_i]\right)\frac{\partial^{c}
\ell_{\alpha}}{\partial\theta^j}\right. \nonumber\\
&&+\left.\ell_{\alpha}\left[[\ell_{-\alpha},\Delta_i],\Delta_j\right]+
\left(\frac{1-\alpha}{2}\right)^{\frac{2\alpha}{1-\alpha}}\ell_{\alpha}
^{\frac{1+\alpha}{1-\alpha}}\left[[\Delta_j,[\ell_{\alpha},\Delta_i]\right]
\right\}\nonumber\\
&=&\frac{2\alpha}{1-\alpha}\mbox{Tr}\left(\frac{\partial^{c}\ell_{-\alpha}}
{\partial\theta^i}\frac{\partial^{c}\ell_{\alpha}}{\partial\theta^j}\right)
\label{scholari}
\end{eqnarray}
Collecting together (\ref{juninho}),(\ref{rivaldo}) and (\ref{scholari}) we
conclude that 
\begin{equation}
\frac{\partial^2\widetilde{\Psi}_{\alpha}(\theta)}
{\partial\theta^i\partial\theta^j}=\mbox{Tr}\left(
\frac{\partial^{c}\ell_{-\alpha}}{\partial\theta^i}
\frac{\partial^{c}\ell_{\alpha}}{\partial\theta^j}\right)
\end{equation}

On the other hand, by the same argument used to find (\ref{juninho}), we have that
the {\em WYD} in this $\widehat{\nabla}^{\alpha}$-affine
coordinate system assumes the form
\begin{equation}
g^{(\alpha)}_{ij}(\theta)=\mbox{Tr}\left(X_j\frac{\partial\ell_{-\alpha}}
{\partial\theta^i}\right)=\mbox{Tr}\left(
\frac{\partial^{c}\ell_{-\alpha}}{\partial\theta^i}
\frac{\partial^{c}\ell_{\alpha}}{\partial\theta^j}\right),
\end{equation}
which proves the first assertion of the lemma. For the second part of the lemma,
we have seen in the previous section that there exists a $\widehat{\nabla}^{-\alpha}$-affine
coordinate system $\xi=\{\xi_1,\ldots,\xi_{n+1}\}$ in terms of which we can write
\[\ell_{-\alpha}=\xi_1Y^1+\cdots+\xi_{n+1}Y^{n+1},\]
for some other set of linearly independent operators $\{Y^1,\ldots,Y^{n+1}\}$. Now
following the same reasoning that led to (\ref{ronaldo}) we obtain that
\begin{eqnarray}
\frac{\partial\widetilde{\Psi}_{\alpha}(\theta)}
{\partial\theta^i}&=&\frac{1-\alpha}{2}\mbox{Tr}\left(\frac{\partial\ell_{\alpha}
\ell_{-\alpha}}{\partial\theta^i}\right)\nonumber\\
&=&\frac{1-\alpha}{2}\mbox{Tr}\left(X_i\ell_{\alpha}+\ell_{\alpha}\frac{\partial
\ell_{-\alpha}}{\partial\theta^i}\right)\nonumber\\
&=&\frac{1-\alpha}{2}\mbox{Tr}\left[\left(1+\frac{1+\alpha}{1-\alpha}\right)
X_i\ell_{-\alpha}\right]\nonumber\\
&=&\mbox{Tr}\left[X_i\left(\xi_1Y^1+\cdots+\xi_{n+1}Y^{n+1}\right)\right]\nonumber\\
&=&\xi_1\mbox{Tr}\left(X_iY^1\right)+\cdots+\xi_{n+1}\mbox{Tr}\left(X_iY^{n+1}\right)\\
&=&\sum_{j=1}^{n+1}\mbox{Tr}\left(X_iY^j\right)\xi_j.
\end{eqnarray}
This means that the coordinate system $(\widetilde{\eta})$ is affinely related to
$(\xi)$ and therefore it is itself $\widehat{\nabla}^{-\alpha}$-affine.

\vspace{0.4in}

We end this section with the next theorem, which is the extension for a general 
$\alpha$-connections of the result proved in
\cite{GrasselliStreater01a} for the case $\alpha=\pm 1$.

\begin{theorem}
For a fixed value of $\alpha\in(-1,1)$, suppose that the connections $\nabla^{(\alpha)}$ 
and $\nabla^{(-\alpha)}$ are dual
with respect to a Riemannian metric $\widehat{g}$ on $\widehat{\cal M}$. Then there
exist a constant (independent of $\sigma$) $(n+1)\times (n+1)$ matrix $M$,
such that $(\widehat{g}_{\sigma})_{ij}={\displaystyle \sum_{k=1}^{n+1}}
M_i^k(\widehat{g}^{(\alpha)}_{\sigma})_{kj}$, in some $\alpha$-affine
coordinate system. \label{unique1}
\end{theorem}
{\em Proof:} Since the two connections are flat on the extend manifold $\widehat{\Man}$, 
theorem~\ref{dualities} tell us that there exist dual coordinate systems $(\theta,\eta)$
such that $\theta$ is $\nabla^{(\alpha)}$-affine and $\eta$ is
$\nabla^{(-\alpha)}$-affine. Using lemma \ref{alphalemma}, we know that 
the function $\widetilde{\Psi}_{\alpha}(\theta)=\frac{2}{1+2}\mbox{Tr}\sigma(\theta)$ satisfies
\begin{equation}
\widehat{g}^{(\alpha)}_{ij}(\theta)=\frac{\partial^2\widetilde{\Psi}_{\alpha}(\theta)}
{\partial\theta^i\partial\theta^j}
\end{equation}
and also that 
\begin{equation}
\widetilde{\eta}_i=\frac{\partial\widetilde{\Psi}_{\alpha}(\theta)}{\partial\theta^i}
\end{equation}
is a another $\widehat{\nabla}^{(-\alpha)}$-affine coordinate system for $\widehat{\Man}$.
Therefore, the coordinate systems $(\eta)$ and $(\widetilde{\eta})$ are related
by an affine transformation, so there must exist a matrix $M$ and numbers $(a_1,\ldots,
a_{n+1})$ such that  
\begin{equation}
\eta_i = \sum_{k=1}^{n+1}M_i^k\widetilde{\eta}_k + a_i.
\label{affine}
\end{equation}

But from theorem~\ref{potentials},
there exists a potential function $\Psi(\theta)$ such that
\[\widehat{g}_{ij}(\theta)=\frac{\partial^2\Psi(\theta)}{\partial\theta^i\partial\theta^j}\]
and
\[\eta_i=\frac{\partial\Psi(\theta)}{\partial\theta^i}.\]
Equation (\ref{affine}) then gives
\[\frac{\partial\Psi(\theta)}{\partial\theta^i} =
\sum_{k=1}^{n+1}M_i^k\frac{\partial\widetilde{\Psi}_{\alpha}(\theta)}{\partial\theta^k} +
a_i,\]
and differentiating this equation  with respect to $\theta^j$ leads to
\begin{equation}
\widehat{g}_{ij}(\theta)=\frac{\partial^2\Psi(\theta)}{\partial\theta^i\partial\theta^j}
=\sum_{k=1}^{n+1}M_i^k
\frac{\partial^2\widetilde{\Psi}_{\alpha}(\theta)}{\partial\theta^j\partial\theta^k}
=\sum_{k=1}^{n+1}M_i^k\widehat{g}^{(\alpha)}_{kj}(\theta).
\label{psis}
\end{equation}

\vspace{0.2in}

\section{The condition of monotonicity}

We have seen in the previous section that requiring duality
between the $\nabla^{(\alpha)}$ and $\nabla^{(-\alpha)}$ connections reduces the set of
possible Riemannian metrics on $\widehat{\cal M}$ to matrix multiples of
the {\em WYD} metric. Following \cite{GrasselliStreater01a}, we now investigate the effect of imposing a
monotonicity property on this set.

Recall that the $-1$-representation is the limiting case $\alpha=-1$ of the $\alpha$-representations defined
in section \ref{alpharepresentation}. If we use it to define
to define a Riemannian metric $\widehat{g}$ on $\widehat{\cal M}$
by means of the inner product
$\langle\cdot,\cdot\rangle_{\rho}$ in ${\cal A}\subset B({\cal H}^N)$,
then we say that $\widehat{g}$ is {\em monotone} if and only if
\begin{equation}
\left\langle S(A^{(-1)}),S(A^{(-1)})\right\rangle_{S(\rho)} \leq \left\langle
A^{(-1)},A^{(-1)}\right\rangle_{\rho},
\label{mono}
\end{equation}
for every $\rho \in {\cal M}$, $A \in T_{\rho}{\cal M}$, and every
completely positive, trace preserving map $S:{\cal A} \rightarrow {\cal
A}$.

For any metric $\widehat{g}$ on $T\widehat{\cal M}$, define the positive (super)
operator $K_\sigma$ on $\cal A$ by
\begin{equation}
\widehat{g}_\sigma(\widehat{A},\widehat{B})=\left\langle
\widehat{A}^{(-1)},K_\sigma\left(\widehat{B}^{(-1)}\right)
\right\rangle_{\scriptscriptstyle H \!
S}=\mbox{Tr}\left(\widehat{A}^{(-1)}K_\sigma
\left(\widehat{B}^{(-1)}\right)\right). \label{metric}
\end{equation}
Note that our $K$ is denoted $K^{-1}$ by Petz in \cite{Petz96}. Define also the
(super) operators, $L_\sigma X:=\sigma X$ and $R_\sigma X:=X\sigma$, for
$X \in {\cal A}$, which are also positive. The aforementioned characterization of monotone metrics obtained 
by Petz is the content of the following theorem.

\begin{theorem}[Petz 96]
A Riemannian metric $g$ on $\cal A$ is monotone if and only if
\[K_{\sigma}=\left(R_{\sigma}^{1/2}f(L_{\sigma}R_{\sigma}^{-1})R_{\sigma}^{1/2}
\right)^{-1},\]
where $K_{\sigma}$ is defined in~(\ref{metric}) and $f:R^{+}\rightarrow R^{+}$
is an operator monotone function satisfying $f(t)=t f(t^{-1}).$
\label{function}
\end{theorem}

In particular, the {\em WYD} metric is monotone and its corresponding operator
monotone function is
\begin{equation}
f_p(x)=\frac{p(1-p)(x-1)^2}{(x^p-1)(x^{1-p}-1)},
\end{equation}
for $p=\frac{1+\alpha}{2}$ \cite{HasegawaPetz96}. 

Combining this characterization with our theorem~(\ref{unique1}), we
obtain the following improved uniqueness result.

\begin{theorem}
If the connections $\nabla^{(\alpha)}$ and $\nabla^{(-\alpha)}$ are dual with respect
to a monotone Riemannian metric $\widehat{g}$ on $\widehat{\cal M}$, then $\widehat{g}$ is a
{\em scalar} multiple of the {\em WYD} metric.
\label{unique2}
\end{theorem}
{\em Proof:} Let $\theta=(\theta^1,\ldots,\theta^n)$ be the $\nabla^{(\alpha)}$-affine coordinate
system of theorem \ref{unique1}. Given $\sigma\in\widehat{\cal M}$, we have that $T_\sigma{\cal M}\simeq \Alg$.
In particular, $\left\{\frac{\partial\sigma}{\partial\theta^1},
\ldots ,\frac{\partial\sigma}{\partial\theta^n}\right\}$ is the basis for $\Alg$ obtained as the
$-1$-representation of $\left\{\frac{\partial}{\partial\theta^1},
\ldots ,\frac{\partial}{\partial\theta^n}\right\}$. Now let $K^g$ and $K^{(\alpha)}$ be the kernels 
of $\widehat{g}$ and $\widehat{g}^{(\alpha)}$, respectively. Then it follows from theorem \ref{unique1} that
\begin{eqnarray}
\left\langle\frac{\partial\sigma}{\partial\theta^i},K^g_\sigma\left(\frac{\partial\sigma}{\partial\theta^j}\right)
\right\rangle_{\scriptscriptstyle H\!S}&=& \widehat{g}_\sigma\left(\frac{\partial}{\partial\theta^i},
\frac{\partial}{\partial\theta^j}\right)
=(\widehat{g}_\sigma)_{ij} \nonumber \\
&=& \sum_{k=1}^{n+1} M_i^k(\widehat{g}^{(\alpha)}_\sigma)_{kj} \nonumber \\
&=& \sum_{k=1}^{n+1} M_i^k \widehat{g}^{(\alpha)}_\sigma\left(\frac{\partial}{\partial\theta^k},
\frac{\partial}{\partial\theta^j}\right) \nonumber \\
&=& \sum_{k=1}^{n+1} M_i^k \left\langle\frac{\partial\sigma}{\partial\theta^k},K^{(\alpha)}_\sigma
\left(\frac{\partial\sigma}{\partial\theta^j}\right)\right\rangle_{\scriptscriptstyle H\!S}.
\end{eqnarray}

Thus, as operators on $\Alg$, the kernels $K^g$ and $K^{(\alpha)}$ are related by
\begin{equation}
K^g_\sigma=M K^{(\alpha)}_\sigma.
\end{equation}
Therefore, if $f^g$ and $f^{(\alpha)}$ are the operator
monotone functions
corresponding respectively to $g$ and $g^{(\alpha)}$, from
theorem~\ref{function}, we have
\begin{eqnarray*}
\left(R_{\sigma}^{1/2}f^g(L_{\sigma}R_{\sigma}^{-1})R_{\sigma}^{1/2}\right)^{-1}
& = & M\left(R_{\sigma}^{1/2}f^{(\alpha)}
(L_{\sigma}R_{\sigma}^{-1})R_{\sigma}^{1/2}\right)^{-1}\\
\left(R_{\sigma}^{1/2}f^g(L_{\sigma}R_{\sigma}^{-1})R_{\sigma}^{1/2}\right)M
& = & \left(R_{\sigma}^{1/2}f^{(\alpha)}
(L_{\sigma}R_{\sigma}^{-1})R_{\sigma}^{1/2}\right)\\ M & = &
f^g(L_{\sigma}R_{\sigma}^{-1})^{-1}f^{(\alpha)}
(L_{\sigma}R_{\sigma}^{-1}),
\end{eqnarray*}
as everything commutes.
Thus, the operator $M$ is given as a function of the operator
$L_{\sigma}R_\sigma^{-1}$, but it is itself independent of the point $\sigma$,
so we conclude that it must be a scalar multiple of the identity
operator.

\section{Discussion}

With the result of this paper, we have completed the programme initiated in \cite{GrasselliStreater01a}
of characterizing the {\em BKM} and the {\em WYD} metrics in terms of the combining requirement of monotonicity
and duality. The monotonicity condition has an appealing motivation coming from
estimation theory. If we interpret the geodesic distance between two density
matrices as a measure of their statistical distinguishability, then 
(\ref{mono}) tells us that they will become less distinguishable if we 
introduce randomness into the system under consideration. In other words, 
their distance decreases under coarse-graining. 

As it is, estimation theory is more basic than physics itself, since it does 
not assume any particular underlying physical process, being just a tool to
help analyze statistical data. Nevertheless, the interpretation above carries
over to statistical mechanical systems as well, where stochastic (i.e 
completely positive, trace-preserving) maps appear as a mathematical 
implementation of the time evolution of a system whose states are described 
by density matrices \cite{Streater95c}. In this case, monotonicity means 
that the distance between different states decreases under the same time
evolution. If it decreases asymptotically to zero for any two points in a 
certain set of `initial' states, then we are in the presence of a fixed point
for the dynamics, or in other words, an equilibrium state. From all this,
it seems that imposing a monotonicity condition on the possible Riemannian
metrics on a statistical manifold is not at all an artificial technicality.

Our motivation behind Amari's duality is less general  
and ultimately rests upon quantum statistical mechanics alone \cite{Streater95a,Streater96}. Recall
that the von Neumann entropy for a state $\rho \in \Man$ is defined as 
\cite{vonNeumann96}
\begin{equation}
S(\rho):=-\mbox{Tr}(\rho \log \rho)
\end{equation}
and that the relative (Kullback-Leibler) entropy of the state $\rho$ 
given the state $\sigma$ is
\begin{equation}
S(\rho|\sigma)=\mbox{Tr}[\rho(\log\rho-\log\sigma)]
\label{Kullback}
\end{equation}
Now let us choose a set of $m\leq n$ observables $Y_1,\ldots,Y_m$ such that
the set $\{{\bf 1},Y_1,\ldots,Y_m\}$ is a basis for ${\cal A}$. 
Among all possible observables in ${\cal A}$, these ones represent the 
{\em slow variables} of the theory, that is, those whose means we can measure
at any given time. Then it is an easy exercise, using the Lagrange multipliers
technique, to show that the states which maximize the von Neumann entropy 
subject to keeping the means of all $\{Y_i\}$, $i=1,\ldots,m$, constant 
are the Gibbs states of the form
\begin{equation}
\rho=\exp\left(\theta^1Y_1 + \cdots + \theta^mY_m -
\Psi{\bf 1}\right),
\end{equation}
where $\Psi(\theta)$ is determined by the normalization condition 
$\mbox{Tr}\rho=1$. For example, if $Y_1=H$ is the energy operator, then 
we obtain the so called canonical ensemble, whereas if we have 
$Y_1=H, Y_2=N$ where $N$ is the number of particles, we get the 
grand canonical ensemble. We immediately recognize these states as
constituting a $\nabla^{(1)}$-flat, $m$-dimensional, submanifold 
$\Stat_m \subset \Stat$, which is determined by our choice of 
$Y_1,\ldots,Y_m$, that is, by our choice of the level of description adopted.
 
Inasmuch as entropy is negative information, the principle of {\em maximum
entropy}, advocated in information theory and statistical physics
by Jaynes \cite{Jaynes57a, Jaynes57b}, tells us that, 
if the only information available about the system under 
consideration are the means of the random variables $Y_1,\ldots,Y_m$, then we
should take as the state of the system the element in $\Stat_m$ with these
means. The replacement of the {\em true} state $\rho \in \Stat$ by the one 
in $\Stat_m$ with the same means for $Y_1,\ldots,Y_m$ is a reflection
of our ignorance of what really goes on with the system. It is the 
least biased choice of state given the information available.

The point of view in statistical dynamics \cite{Streater95c} is somewhat 
different, in the sense that it regards the same {\em replacement} as part of
the true dynamics of the system. For instance, the heat transfer in a local 
region of a fluid happens $10^8$ times faster then most chemical reactions
\cite{Fowler80}, so we can choose to regard the concentrations of the 
chemicals reacting as the slow variables while all other observables are
thermalized (maximum entropy) along each time step in the dynamics. The skill
of the scientist using statistical dynamics thus resides in correctly 
identifying which are the slow variables of the problem at hand and then 
following the time evolution of the system, which involves, apart from
a stochastic dynamics particular to each problem, successive {\em projections}
onto $\Stat_m$.

Information geometry provides a mathematical meaning for this projection
\cite{Balian86,Streater95a}.
It is well known that the relative entropy (\ref{Kullback}) is the 
statistical divergence associated with the dualistic triple 
$(g^{\scriptscriptstyle B},\nabla^{(1)},\nabla^{(-1)})$ \cite{Nagaoka95}. 
It then follows from
the general theory \cite{AmariNagaoka00} that, given an arbitrary
point $\rho \in \Stat$, the point in $\Stat_m$ (which is 
$\nabla^{(1)}$-flat) that minimizes $S(\rho|\sigma)$ is obtained uniquely 
by following a $-1$-geodesic from $\rho$ that intercepts $\Stat$ orthogonally
with respect to the {\em BKM} metric $g^{\scriptscriptstyle B}$. 
This is equivalent
to the projection described above (maximum entropy subject to constant means)
precisely because a path preserving the mean parameters (or mixture 
coordinates) is a $-1$-geodesic, that is, a straight line for the 
mixture connection.

However, if $g$ is a general monotone metric, with respect to which 
$\nabla^{(1)}$ and $\nabla^{(-1)}$ are not necessarily dual, then the 
relative entropy might fail to be a divergence for 
$(g,\nabla^{(1)},\nabla^{(-1)})$ and nothing 
guarantees that minimizing $S(\rho|\sigma)$ will produce a point in
$\Stat_m$ connected to $\rho$ by a $-1$-geodesic intersecting $\Stat_m$
perpendicularly with respect to $g$. Information geometry no longer  
provides a mathematical implementation for statistical dynamics anymore.

As a final word for this paper, let us mention that a corollary to theorem \ref{unique2} is the fact
that the relation (\ref{convex}) {\em does not} hold for the quantum $\alpha$--connections defined 
using the $\alpha$--representations as in section \ref{alphaconn}. If it did, a simple calculation shows that
$\nabla^{(\alpha})$ and $\nabla^{(-\alpha)}$ would then be dual with respect to the {\em BKM} metric (since 
the $\pm 1$--connections are). But from theorem \ref{unique2}, this would imply that the {\em BKM} is
a scalar multiple of the {\em WYD}, which is only true in the extreme cases $\alpha=\pm 1$.

\vspace{0.3in}
\noindent
{\bf Acknowledgements:} The result of this paper was first announced during the conference Information Geometry and its Applications,
held in Pescara, July 2002. I would like to thank the local organizers, especially P. Gibilisco, for their display of 
Italian hospitality. I also thank A. Jen\v{c}ov\'{a}, H. Hasegawa and R.F. Streater for their comments and suggestions.   

\def\cprime{$'$}

\end{document}